\documentclass[%
 reprint,
 amsmath,amssymb,
 aps,usenames,dvipsnames,nofootinbib
]{revtex4-1}

\usepackage{xcolor}
\usepackage{graphicx}% Include figure files
\usepackage{dcolumn}% Align table columns on decimal point
\usepackage{bm}% bold math
\usepackage{color}
\usepackage{ulem,units}

\newcommand{\SPA}{School of Physics and Astronomy, Monash University, Vic 3800, Australia}
\newcommand{\OzGravMonash}{OzGrav: The ARC Centre of Excellence for Gravitational Wave Discovery, Clayton VIC 3800, Australia}

\begin{document}
\preprint{APS/123-QED}
\title{Accelerated detection of the binary neutron star gravitational-wave background}

\author{Francisco Hernandez Vivanco}
\email{francisco.hernandezvivanco@monash.edu}
\affiliation{\SPA}
\affiliation{\OzGravMonash}

\author{Rory Smith}
\email{rory.smith@monash.edu}
\author{Eric Thrane}
\author{Paul D. Lasky}
\affiliation{\SPA}
\affiliation{\OzGravMonash}

\begin{abstract}
    Most gravitational-wave signals from binary neutron star coalescences are too weak to be individually resolved with current detectors.
    We demonstrate how to extract a population of sub-threshold binary neutron star signals using Bayesian parameter estimation.
    Assuming a merger rate of one signal every two hours, we find that this gravitational-wave background can be detected after approximately three months of observation with Advanced LIGO and Virgo at design sensitivity, versus several years using the standard cross-correlation algorithm.
    We show that the algorithm can distinguish different neutron star equations of state using roughly seven months of Advanced LIGO and Virgo design-sensitivity data.
    This is in contrast to the standard cross-correlation method, which cannot.
\end{abstract}

\maketitle

\section{Introduction} \label{Introduction}
The average time between binary neutron star coalescences somewhere in the universe is currently constrained to be $\unit[13^{+49}_{-9}]{s}$~\cite{LSC_stochastic_paper:2018,LSC_GW170817}.
It is expected that the advanced gravitational-wave detector network will only be able to resolve individual binary neutron star mergers out to an average distance of $\unit[190]{Mpc}$~\cite{aligo:2015},  which corresponds to just $\approx 0.001$~\% of the~ $\approx 2\times10^6$ binary neutron star mergers that take place somewhere in the Universe each year.
These distant, unresolved signals form an astrophysical gravitational-wave background in Advanced LIGO (aLIGO)~\cite{Aasi:2015} and Virgo~\cite{Acernese:2014}, which has been a longtime target for advanced gravitational-wave observatories~\cite{Allen:1999,Allen:1996}.

Recently, Smith and Thrane \cite{Smith:2017vfk} developed an optimal method for detecting astrophysical gravitational-wave backgrounds from unresolved binary merger signals. The search applies Bayesian parameter estimation to all available data to compute the posterior probability distribution on the ``duty cycle'': the fraction of all data that contains a gravitational signal from compact binary mergers.
In Ref. \cite{Smith:2017vfk}, it was shown how the gravitational-wave background from unresolved binary black holes could be detected with around one day of design-sensitivity aLIGO data versus several years for the standard cross-correlation search. The optimal search from~\cite{Smith:2017vfk} is much more effective than the cross-correlation search
because it employs a likelihood function which describes the properties of both astrophysical gravitational-wave signals and noise features.

One of the simplifying assumptions in~\cite{Smith:2017vfk}, is that there is only one gravitational-wave signal in any given $\unit[4]{s}$ segment of data.
This is an excellent approximation for relatively high mass binary black hole signals above $\approx\unit[20]{Hz}$.
Each signal is short (typically $\lesssim\unit[1]{s}$) and the probability of getting two mergers in one segment is $\approx 10^{-4}$~\cite{PhysRevD.84.084004,PhysRevD.87.043009}.
However, binary neutron stars are different.
Binary neutron star signals can last for $\approx\unit[100]{s}$ for frequencies above $\unit[20]{Hz}$~\cite{PhysRevD.94.044031} with typically 15 unresolved binary neutron star signals at any given time~\cite{LSC_stochastic_paper:2018}.
In this paper, we extend the optimal method from~\cite{Smith:2017vfk} to measure the binary neutron star background by analyzing a subset of the observing band $f>\unit[100]{Hz}$.
In this band, the binary neutron star signals are shorter and less likely to overlap, and the detection problem is more similar to the binary black hole background.

By excluding data from $\unit[20-100]{Hz}$, the search is less sensitive than it could be in theory, but it is still substantially more sensitive than the standard cross-correlation search.
We demonstrate that our search strategy is capable of detecting the binary neutron star background with around three months of design sensitivity aLIGO and Virgo data, versus several years using cross-correlation.
Furthermore, we show that the binary neutron star background encodes information about the neutron star equation of state in the form of tidal effects, which affect the phase evolution of the gravitational waveform at high frequencies $\gtrsim\unit[100]{Hz}$~\cite{Harry:2018hke}.
Even though the binaries that make up the background are unresolved, analysis of many sub-threshold signals can be combined to ascertain shared properties.
In this work, we consider a signal to be subthreshold if the matched-filter signal-to-noise-ratio~$<$12. All signals with signal-to-noise-ratio~$\geq 12$ are removed.
We show that we can distinguish between different equations of state with around seven months of aLIGO and Virgo data.
This is remarkable since the cross-correlation approach probably cannot distinguish between a $30 M_\odot$ binary black hole background from a $1.4 M_\odot$  binary neutron star background \cite{PhysRevX.6.031018}.

This paper is organized as follows. In Sec.~\ref{sec:Method} we describe the framework of our search. Then, in Sec \ref{sec:time_to_detection} we estimate the time-to-detection of the binary neutron star background assuming realistic merger rates while in Sec.~\ref{sec:EoS} we show that our search can be used to constrain the equation of state. We conclude in Sec.~\ref{sec:Conclusions}.

\section{Method} \label{sec:Method}
We follow~\cite{Smith:2017vfk}, which considered an optimal search for a population of weak gravitational waves from binary black holes. 
However, there are some subtleties that arise when the technique is applied to binary neutron stars.
In the binary black hole analysis, the data are divided into $\unit[4]{s}$ segments.
The likelihood for the data in segment $i$ can be written as
\begin{equation} \label{eq:generalized_likelihood}
    \mathcal{L}(s_i|\theta,\xi) = \xi \mathcal{L}(s_i,\theta) + (1-\xi)\mathcal{L}(s_i,\varnothing) \,.
\end{equation}

Here $\mathcal{L}(s_i,\theta$) and  $\mathcal{L}(s_i,\varnothing)$ are the likelihood functions assuming the signal and noise hypothesis and $\theta$ represents the binary neutron star parameters. The hyper-parameter $\xi$ is referred as the ``duty cycle.''
For one segment, it is the probability that the segment contains a signal.
When multiple segments are combined, $\xi$ can be interpreted as the fraction of all data segments that contain a signal~\cite{Smith:2017vfk}.
Since we are interested in the hyper-parameter $\xi$, we can marginalize over the binary parameters $\theta$ to obtain the posterior distribution of $\xi$,
\begin{equation} \label{eq:evidence_likelihood}
    \mathcal{L}(s_i|\xi) = \xi \mathcal{Z}_s + (1-\xi)\mathcal{Z}_n \,.
\end{equation}
where $\mathcal{Z}_s$ and $\mathcal{Z}_n$ are the signal and noise evidences. We then combine data from $N$ different segments,

\begin{align}
    {\cal L}(\vec{s}|\xi) = \prod_i^N {\cal L}(s_i|\xi) .
\end{align}
The joint likelihood can be used to construct a posterior for $\xi$.
A binary black hole background is detected when the credible interval for $\xi$ excludes zero.
The analysis is made relatively simple due to the simple form of Eq.~\ref{eq:generalized_likelihood}, which we obtain by assuming that each segment contains at most one binary merger.
While this is a great approximation for binary black holes in the LIGO/Virgo band, the situation is different for binary neutron stars.

\begin{figure}[!t]
    \centering
    \includegraphics[width=\columnwidth]{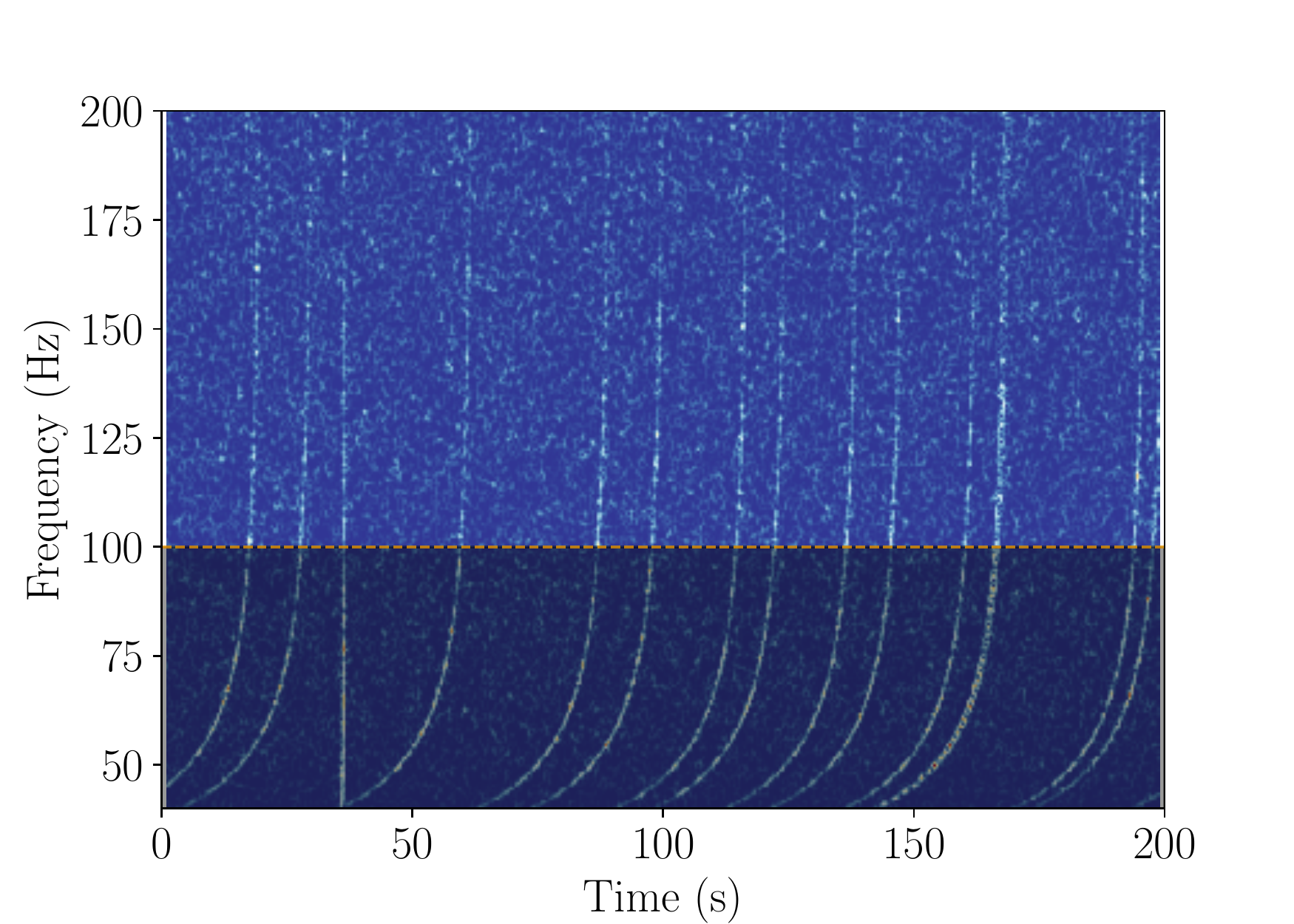}
    \caption{A spectrogram of $\unit[200]{s}$ of simulated data showing thirteen binary neutron stars and one binary black hole merger (the black hole merger is the third signal from left to right). Our strategy for detecting the binary neutron star background is based around searching for signals above a low-frequency cutoff $f_{\text{low}}=100$ Hz (red dotted line), where binary neutron star signals can be cleanly separated into data segments on the order of $\unit[16]{s}$ in duration, even though the signals overlap at lower frequencies }
    \label{fig:spectrogam}
\end{figure}

In Fig.~\ref{fig:spectrogam}, we plot a typical spectrogram  showing $\unit[200]{s}$ of simulated gravitational-wave strain from binary neutron stars.
For illustrative purposes, we have made the strain large enough to see the binary neutron star signals by eye. In reality, they would contribute to a binary neutron star background composed of unresolved signals, too weak to detect individually.
The figure illustrates how the simplifying assumption from our binary black hole analysis does not apply to binary neutron stars: the typical binary black hole duration is long compared to the typical time between binary neutron stars, and so there are likely to be many (typically 15) binary neutron stars in the observing band at one time.

There are different ways to deal with this added complexity.
One method, which would preserve the optimality of the search, would be to fit for many binary neutron star signals simultaneously in one long segment.
This would introduce an additional fifteen parameters per binary neutron star.
In theory, a reversible jump Monte Carlo method could be employed in order to take into account the unknown number of binary neutron stars in each segment~\cite{Green:1995}.
The clear disadvantage of this approach is that it is likely to  increase dramatically the computational cost of the search because the dimensionality of the evidence integral would become too large.

Another option, left for future exploration, is to change the concept of a ``segment'' from a span of time that contains a binary neutron star signal to a span of time which contains a binary neutron-star merger.
If we recast segments in this way, we can split the data into short, $\unit[200]{ms}$ segments.
On the one hand, this would likely increase the computational cost of the search by increasing the number of segments.
On the other hand, it would allow us to extend $f_\text{low}$ all the way down to $\unit[10]{Hz}$, improving the search sensitivity beyond what is documented here.

An alternative approach is to design an optimal search in a limited observing band in order to control the computational cost.
To do this, we impose a low-frequency cutoff $f_{\text{low}}=100$ Hz, and only analyze strain data at frequencies above $f_{\text{low}}$. 
Employing this cut-off enables us to apply our previously developed framework, while still achieving a substantial improvement in sensitivity compared to cross-correlation.

Above $100$Hz, we only observe the last $\approx \unit[4]{s}$ of the binary neutron star signal, which reduces the cost of data analysis
(the computational cost scales linearly with the signal duration).
Finally, we note that the vast majority of the information about the neutron star equation of state  is contained in frequencies above $100$ Hz (e.g. \cite{Harry:2018hke}).
Below, we demonstrate that we can use the binary neutron star background to infer information about the equation of state.

Now that we have shown how the binary neutron star background can be analyzed analogously to the binary black hole background by focusing on frequencies~$>\unit[100]{Hz}$, we provide additional details for the binary neutron star analysis.
We work with $\unit[16]{s}$ chunks of strain data.
There are two reasons for this choice: it ensures data can be analyzed efficiently and reduces the probability of obtaining an incomplete signal within each data segment.
For each segment $s_i$, we calculate the likelihood function conditional to the signal and noise hypothesis, $\log [\mathcal{L}(s_i|\theta_i)] $ and $\log [\mathcal{L}(s_i|0)] $,
\begin{equation}
    \log [\mathcal{L}(s_i|\theta_i)] \propto -\frac{1}{2} \langle s_i - h(\theta_i) | s_i - h(\theta_i) \rangle \,,
\end{equation}
\begin{equation}
    \log [\mathcal{L}(s_i|0)] \propto \langle s_i|s_i \rangle \,.
\end{equation}
Here, we denote the gravitational waveform in the frequency domain as $h(\theta_i)$. In the results presented in this paper, we generate $h(\theta_i)$ with the \texttt{IMRPhenomD\_NRTidal} waveform approximant \cite{Husa:2016,Khan:2016,Dietrich:2017}, which simulates the merger of a binary neutron star including tidal effects.
We also calculate the usual inner product $\langle a|b \rangle$ starting at the cutoff frequency $f_{\text{low}}=100$Hz,
\begin{equation} \label{inner_product}
    \langle a|b \rangle = 4 \Re \int_{f_{\text{low}}}^{2048\text{Hz}} df \frac{\tilde{a}^*(f) \tilde{b}(f) }{S_n (f)} \,,
\end{equation}
where $S_n$ is the noise power spectral density of the detector. All calculations are performed within the Bayesian inference library \texttt{Bilby}~\cite{Ashton:2018jfp} using the nested sampling algorithm \texttt{Dynesty}~\cite{dynesty}. Finally, the posterior probability distribution on the duty cycle is given by Eq. \ref{eq:xi_posterior} if we assume a uniform prior distribution on $\xi$,
\begin{equation}
    p(\xi | \theta) \propto p(\theta | \xi) \,.
    \label{eq:xi_posterior}
\end{equation}

\subsection{Detection Statistic}
To detect the gravitational-wave background, we calculate the Bayes Factor (BF) which is the ratio between the hypothesis that assumes that the data contains gravitational-wave signals and the null hypothesis (there is no gravitational-wave background),
\begin{equation}
    \textnormal{BF} = \mathcal{Z}_{\textnormal{stoch}} / \mathcal{Z}_{\textnormal{noise}} \,.
\end{equation}
The signal evidence hypothesis, $\mathcal{Z}_{\textnormal{stoch}}$, is obtained by marginalizing over $\xi$ in Eq.
\ref{eq:evidence_likelihood},
\begin{equation}
    \mathcal{Z}_{\textnormal{stoch}} = \int d \xi \mathcal{L}(s_i|\xi)\pi (\xi) \,,
    \label{eq:xi_signal_evidence}
\end{equation}
and the noise evidence hypothesis is simply
\begin{equation}
    \mathcal{Z}_{\textnormal{noise}} = \mathcal{L}(s_i|\xi=0) \,.
    \label{eq:xi_noise_evidence}
\end{equation}
The detection threshold we use is adopted from the convention suggested in \cite{Jeffreys:1961}, where we claim that the binary neutron star gravitational-wave background is detected if $\log \textnormal{BF} = 8 $. 

\begin{figure}[!t]
  
    \includegraphics[width=\columnwidth]{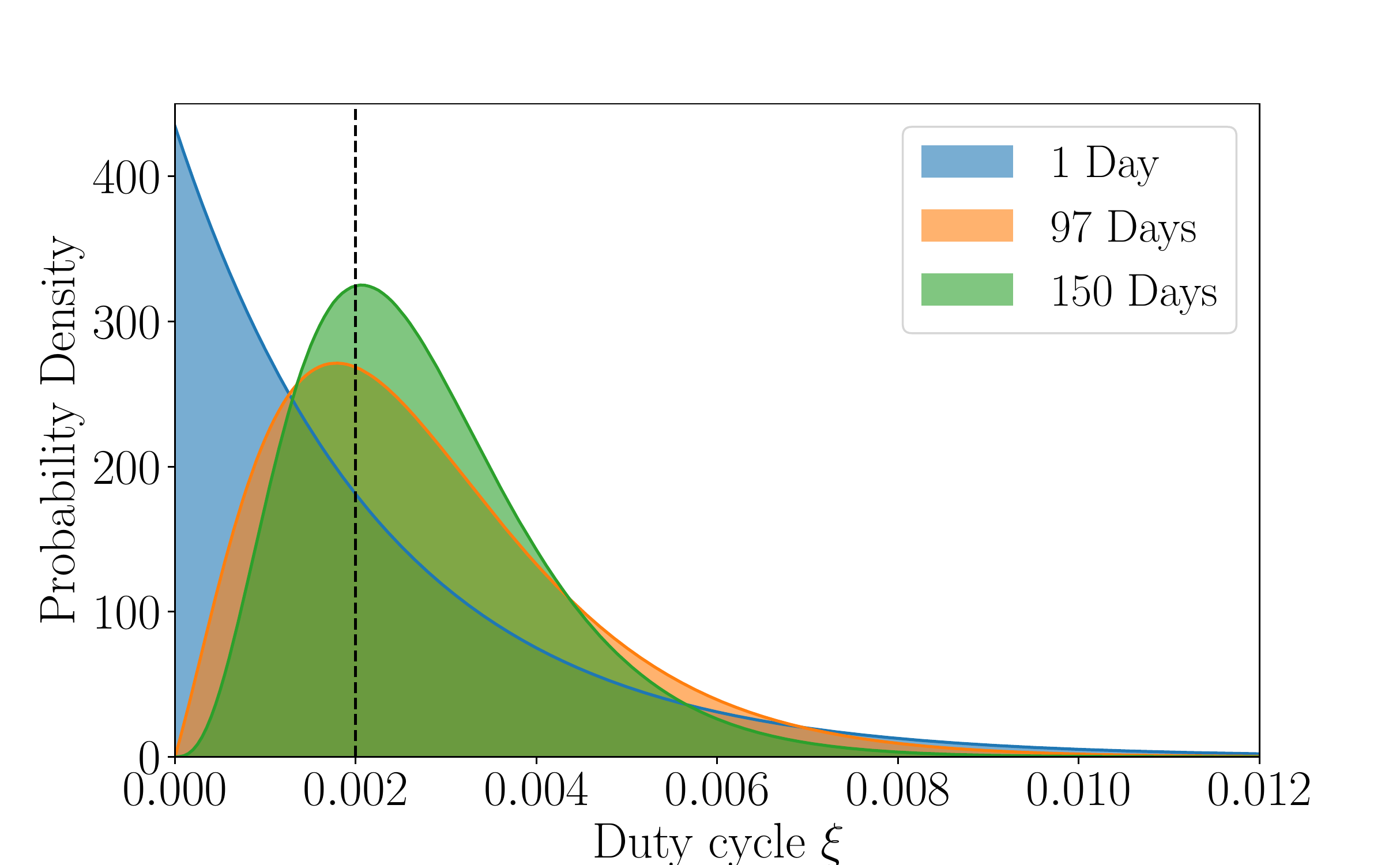}
    \includegraphics[width=1\linewidth]{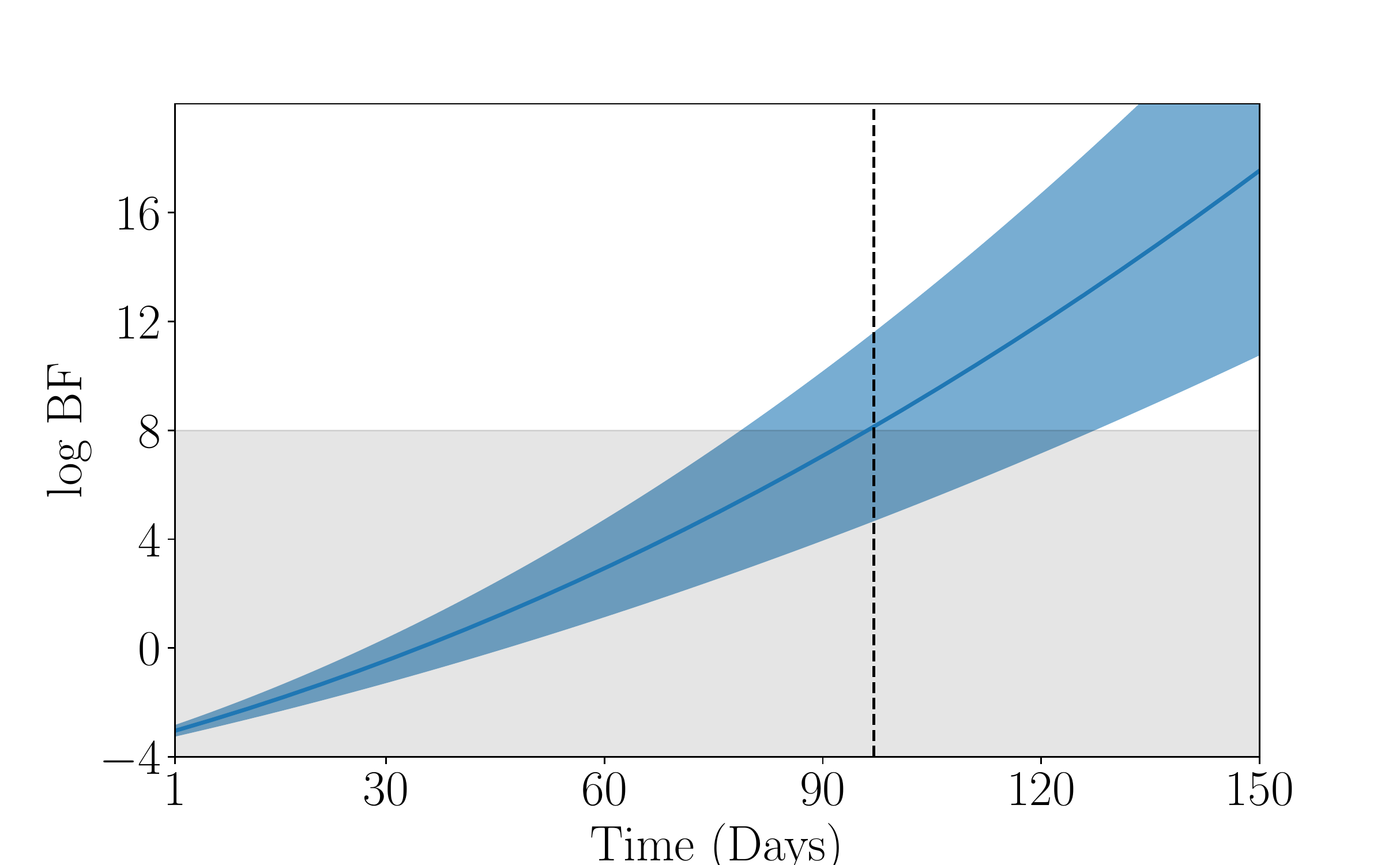}
    \caption{Time to detection of a binary neutron star background.  (Top panel) Posterior distributions for the binary neutron star duty cycle $\xi$ recover the true value.  We simulate aLIGO and Virgo data with a realistic binary neutron star duty cycle of $\xi=0.002$, which corresponds to observing all sub-threshold events in the Universe out to a maximum luminosity distance of 1 Gpc.  The three posterior distributions show the duty-cycle at three different observation times, showing that this eventually becomes unbiased for long observation times.  (Bottom panel).  Evolution of the log Bayes factor showing we can detect the binary neutron star gravitational-wave background after $97^{+32}_{-17}$ days (one-sigma confidence), assuming a merger rate of one signal every two hours.}
    \label{fig:duty_cycle}
\end{figure}

\section{Time-to-detection} \label{sec:time_to_detection}
To obtain an estimate of the observation time before we obtain a detection, we simulate eight hours of data containing signal and noise segments.
The prior distributions used in our search are shown in Tab.~\ref{tab:prior}.
We set the spins to zero in all injections and we also marginalize over time of coalescence, phase of coalescence and luminosity distance. Then, for each segment, we calculate the values of $\mathcal{Z}_s$ and $\mathcal{Z}_n$\footnote{We find that the values of $\mathcal{Z}_s$ are sometimes slightly underestimated using the nested sampling algorithm \texttt{Dynesty}~\cite{dynesty}. However, we show that the posterior distribution of the duty cycle is still unbiased (see Fig. \ref{fig:duty_cycle}) }. In this study, we recover the signal and noise evidence using the same prior as the injected signals distributions. If the priors used for parameter estimation do not match the astrophysical distributions, we find the the duty cycle posterior is unbiased as long as the mass priors overlap. If the mass priors do not overlap, then the search only detects the signals with masses common to both priors.

\begin{table}[t!]%[H] add [H] placement to break table across pages
 \begin{ruledtabular}
 \begin{tabular}{ccccc}
Parameter             &Unit                   & Prior     & Minimum & Maximum \\
 \hline
$\mathcal{M}$         &$M_\odot$              & uniform   & 1.2     & 1.3     \\
$q$                   &-                      & uniform   & 0.5     & 1       \\
RA                    &rad.                   & uniform   & 0       & $2\pi$  \\
DEC                   &rad.                   & cos       & $-\pi/2$& $\pi/2$ \\
$\cos(\theta_{jn})$   &-                      & uniform   & -1      & 1       \\
$\psi$                &rad.                   & uniform   & 0       & $\pi$   \\
$\phi$                &rad.                   & uniform   & 0       & $2\pi$  \\
$d_L$                 &Gpc                    & comoving  & 0.2     & 1       \\
 \end{tabular}
 \end{ruledtabular}
 \caption{Prior distributions used in our search. We simulate and recover the signal evidences with the same distribution of priors. Here, $\mathcal{M}$ is the chirp mass, $q$ is the mass ratio, RA and DEC are the right ascension and declination angles, $\theta_{jn}$ is the inclination angle, $\psi$ is the polarization angle, $\phi$ is the coalescence phase and $d_L$ is the luminosity distance. The comoving prior means that we assume a uniform prior in comoving volume. The dimensionless tidal deformabilities $\Lambda_1$ and $\Lambda_2$ are fixed by the SLY equation of state.}
  \label{tab:prior}
 \end{table}

The choice of the maximum luminosity distance of our search is $\unit[1]{Gpc}$.
Beyond this distance, the search begins to suffer from bias due to underestimate of the signal evidence\footnote{By employing a cut on the maximum distance $d_\text{max}$, we introduce a small bias from the signal coming from $d\gtrsim d_\text{max}$. This bias can be estimated and removed in post-processing using mock data.}.
Work is ongoing to mitigate this bias, which would enable us to extend the maximum distance.
Up to this distance, we expect on average $\approx$ one binary neutron star merger every two hours, which translates to a duty cycle of $\xi=0.002$. 
Due to computational limitations, we cannot simulate the number of segments that will lead to such a duty cycle, however, we can extrapolate our results using a Gaussian mixture model to fit the distributions of the signal and noise evidence \cite{scikit-learn,gmm:2009}.
Our analysis assumes a three detector network -- LIGO Hanford, LIGO Livingston and Virgo -- operating at design sensitivity.

Assuming a merger rate of one signal every two hours, we simulate 30 realizations of the gravitational-wave background and we find that we need to observe for $97^{+32}_{-17}$ days (one-sigma confidence) to reach the detection threshold of  $\log \textnormal{BF} = 8$.
The number of days needed to make a detection would likely decrease if the maximum luminosity distance of the search can be increased. 

The evolution of the posterior probability distribution of $\xi$ as a function of time is presented in Fig. \ref{fig:duty_cycle}.  The result shown in the top panel of Fig. \ref{fig:duty_cycle} shows how the duty cycle posterior distribution evolves over time. After 97 days of observation the posterior recovers the duty-cycle, at this point log BF = 8. After 150 days the $\log \textnormal{BF} \approx 16$.

The strength of the stochastic background signal depends in part on the (somewhat arbitrary) definition of resolvability, which we set to $\rho_\text{th}=12$. We vary $\rho_\text{th}$ to see how the search sensitivity scales with this definition. By lowering $\rho_\text{th}$ to eight, we find that the duty-cycle posterior distribution remains unbiased. However, the time to detection increases from $\sim\unit[3]{months}$ to $\sim\unit[8]{months}$. For $\rho_\text{th}=10$, we find that the background can be detected after $\sim \unit[4]{months}$. Thus, roughly half of the stochastic signal is obtained from signals with signal-to-noise-ratio between eight and ten. It is possible that we might be able to get more stochastic signal from low signal-to-noise ratio events by improving the sampling algorithm and extending the maximum luminosity distance of the search. This is something we hope to explore in future work.

We compare the sensitivity of our Bayesian method to cross-correlation, which is typically cast within the frequentist framework. It is not entirely straightforward to compare frequentist and Bayesian detection statistics. However, if we assume the conventional detection thresholds of $\log(\text{BF})=8$ for the Bayesian algorithm and the “five-sigma” threshold for the cross-correlation analysis, then we find that the Bayesian analysis here will result in a detection approximately ten times faster than the cross-correlation statistic.

For simplicity, in this study we do not account for ``glitches'': noise transients caused by the instrument or the environment, usually from unknown origin~\cite{PhysRevD.98.084016}. 
However, in Ref. \cite{Smith:2017vfk} it was shown that glitches can be included in our search's framework. 

\section{Constraining the equation of state} \label{sec:EoS}
Neutron stars are ideal objects to study matter at supernuclear densities. The behaviour of neutron stars is governed by the equation of state (EOS), which describes a relationship between state variables such as density and pressure (see, e.g., \cite{BALDO2016203,Hebeler:2013,Haensel:2007yy} for a review). This relation, together with the general relativity equations for hydrostatic equilibrium, give a relationship between mass and radius of a neutron star. Such a relation can be constrained by gravitational waves if we can measure the dimensionless tidal deformability $\Lambda$, which determines how much a neutron star deforms in the presence of another massive object. The dimensionless tidal deformability depends on the radius of a neutron star $R$ and the second love number $k_2$, $\Lambda = (2/3) k_2 C^{-5}$~, where $C=GM/(c^2R)$ is the compactness \cite{Flanagan:2008}. Both of these quantities depend on the equation of state and it is for this reason that in principle, constraining the value of $\Lambda$ leads to a constraint on the equation of state. 

Lackey and Wade~\cite{Lackey:2015} showed that the tidal deformability $\Lambda$ can be measured to $\approx$ 10 - 50 \% (depending on the equation of state and mass distribution) with $\approx$40 binary neutron star detections, although the majority of information comes from the loudest $\approx$ five events.
To date, constraints from LIGO place $\Lambda_{1.4}=190^{+390}_{-120 }$ at the $90 \%$ level (for a 1.4~$M_{\odot}$ neutron star)~\cite{LSC_EoS:2018}.

With our method, we can constrain the equation of state using solely sub-threshold signals.
The main idea is that the signal and noise evidences, ${\mathcal Z}_s$ and ${\mathcal Z}_n$ respectively, will be different depending on which equation of state we use as a prior.
We can therefore extract information about the equation of state by calculating the Bayes factor between any two different models,
\begin{equation}
    \textnormal{BF} = \frac{\mathcal{Z}_{\textnormal{EOS}_1 } }{\mathcal{Z}_{\textnormal{EOS}_2}  } \,,
    \label{eq:bayes_factor}
\end{equation}
where $\mathcal{Z}_{\textnormal{EOS}}$ are the Bayesian evidences from Eq. \ref{eq:xi_signal_evidence} with a fixed equation of state prior,
\begin{equation}
    \mathcal{Z}_{\textnormal{EOS}} = \int d\xi \mathcal{L}(s|\xi,\textnormal{EOS})\pi (\xi) \,.
    \label{eq:eos_evidence}
\end{equation}

\begin{figure}[!t]
    \centering
    \includegraphics[width=\columnwidth]{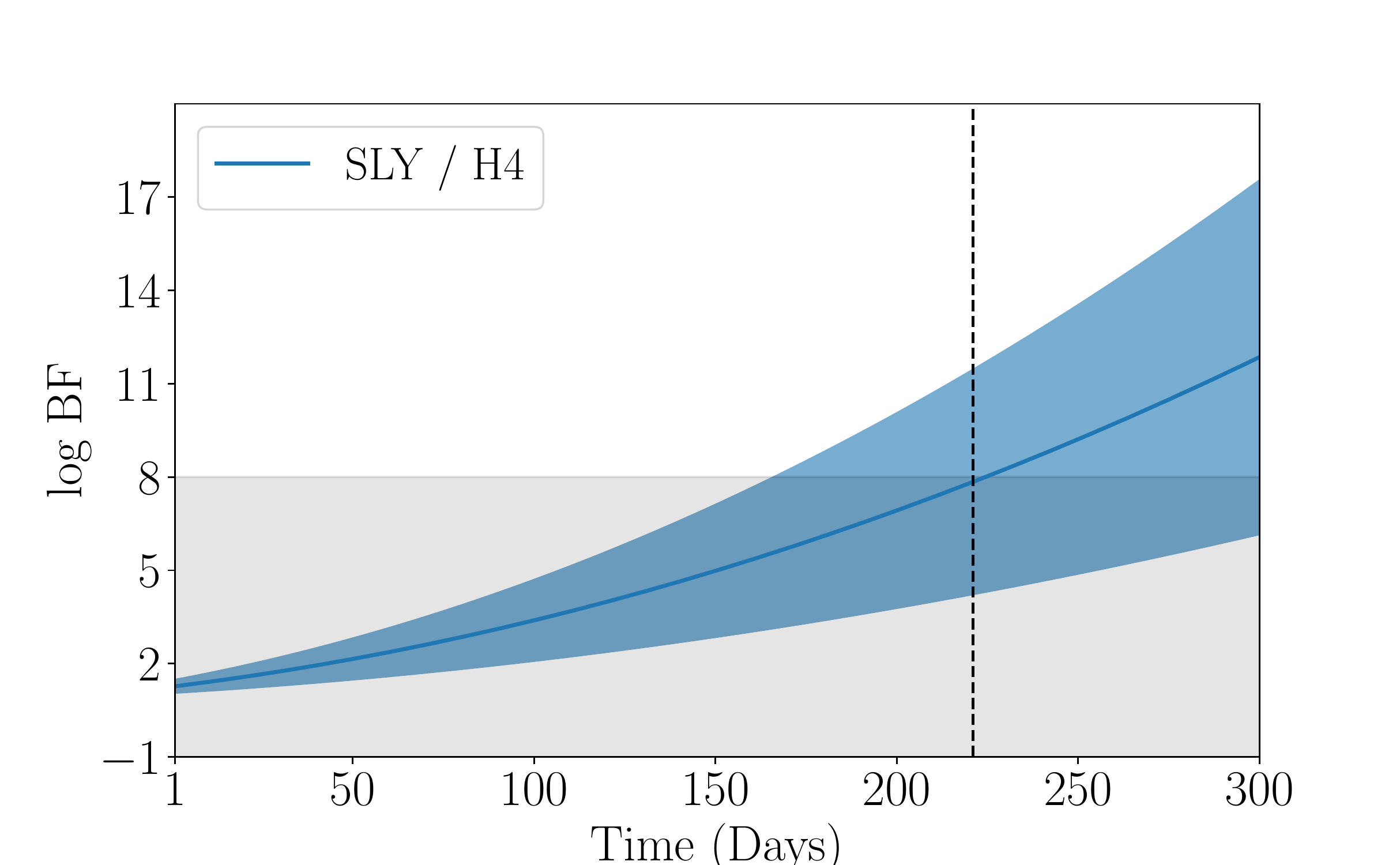}
    \caption{Evolution of the average log BF as a function of time comparing two different equations of state, SLY and H4, for 30 different realizations of the background. The black dotted line is the detection threshold log(BF) = 8, which is reached after $221^{+140}_{-51}$ days of observation. This threshold is reached using exclusively sub-threshold signals.}
    \label{fig:sly_h4}
\end{figure}

Using simulated data, we compare two equations of state covering different values of stiffness: SLY ($\Lambda_{1.4} = 264.5$)~\cite{Douchin:2001sv} and H4 ($\Lambda_{1.4} = 1051.8$)~\cite{Read:2009}.
Here, the sub-index $\Lambda_{1.4}$ stands for the dimentionless tidal deformability of a neutron star of 1.4 solar masses.  We inject simulated data with the SLY equation of state and then recover the evidences $Z_{\text{SLY}}$ and $Z_{\text{H4}}$. We calculate Eq. \ref{eq:xi_posterior}, assuming different observation times starting from one day to 150 days. We then simulate 30 realization of the background and we find that in order to be able to distinguish between SLY and H4, we need to analyze $221^{+140}_{-51}$ days of data (one-sigma confidence), as shown in Fig. \ref{fig:sly_h4}. If we analyze all of the events in our dataset, not just the sub-threshold ones, then we expect  (on average) to distinguish  between these two equations of state after ~1.6 months of observation.

The measurement of GW170817 already rules out H4 at 90\% confidence \cite{LSC_EoS:2018}.
We will learn more about the equation state from the loud, resolvable events.
However, it is remarkable that we can use unresolved binary neutron stars to say anything about the neutron star equation of state.
The constraints obtained by sub-threshold signals can be compared to the results obtained using resolved binary neutron stars as a sanity check, or the two results can be combined to obtain improved sensitivity.

\section{Conclusions} \label{sec:Conclusions}
We derive a method which would detect the binary neutron star gravitational-wave background after $97^{+32}_{-17}$~days of observation using design-sensitivity aLIGO and Virgo data assuming a merger rate of one signal every two hours, which is consistent with a realistic binary neutron star merger rate~\cite{LSC_GW170817}. 
Future studies will focus on optimizing our search to be able to recover the signal evidence at distances larger than 1 Gpc. More sophisticated analyses will additionally include hierarchical Bayesian inference, which can provide new and complementary insights to the distribution of masses and tidal deformabilities of neutron stars \cite{Talbot:2018}. 

\section*{Acknowledgments}
This work is supported through Australian Research Council (ARC) Centre of Excellence CE170100004. ET is supported through ARC Future Fellowship FT150100281.  PDL is supported through ARC Future Fellowship FT160100112 and ARC Discovery Project DP180103155. FHV is supported through the Monash Graduate Scholarship (MGS). The results presented in this manuscript were calculated using the supercomputer clusters at California Institute of Technology and Swinburne University of Technology (OzSTAR). The authors are grateful for computational resources provided by the LIGO Laboratory and supported by National Science Foundation Grants
PHY-0757058 and PHY-0823459.

\bibliography{bibliography}
\end{document}